\documentclass{article}

\usepackage{spconf,amsmath,graphicx,booktabs}
\usepackage{multirow}
\usepackage{makecell}
\usepackage{wrapfig}
\usepackage{hyperref}
\usepackage{tikzpagenodes}
\usepackage{subcaption}
\makeatletter
\newcommand{\manuallabel}[2]{\def\@currentlabel{#2}\label{#1}}
\makeatother
\title{Adaptive Local Implicit Image Function for Arbitrary-scale Super-resolution}
\name{Hongwei Li$^1$, Tao Dai$^2$, Yiming Li$^1$, Xueyi Zou$^3$, Shu-Tao Xia$^1$
\thanks{This work is supported in part by the National Natural Science Foundation of China under Grant 62171248,  the Natural Science Foundation of Guangdong Province 2021A1515011807. Corresponding author: Tao Dai.
}
}
\address{$^{1}$Tsinghua Shenzhen International Graduate School, Tsinghua University \\
$^2$College of Computer Science and Software Engineering, Shenzhen University \\
$^3$Huawei Noah’s Ark Lab\\
\texttt{\{lhw20,li-ym18\}@mails.tsinghua.edu.cn}; \texttt{daitao.edu@gmail.com}\\
\texttt{zouxueyi@huawei.com}; \texttt{xiast@sz.tsinghua.edu.cn}
}
\begin{document}
%\ninept
%
\maketitle
\begin{abstract}
% Recently, implicit neural representations have been successfully progressed in many domains. Pixel values in the image domain can be successfully inferred from neural network models in the continuous space domain. The advantage of introducing NeRF into the super-resolution domain is that multi-scale reduction can be achieved by training a single model, effectively solving the pain point of commercializing super-resolution algorithms to the ground. In recent work (e.g., LIIF), it has been shown that the method can achieve good performance in super-resolution tasks at any scale, but often suffers from high-frequency texture and image structure errors on large magnification restorations. In this work, we propose A-LIIF, a new network design using adaptive implicit image functions for each pixel point, which uses a combination of predictive networks and Basis to adaptively express Pixel-Wise implicit image functions to resolve high-frequency texture and image structure errors. We have conducted extensive experimental and ablation studies, and our A-LIIF achieves state-of-the-art results compared to previous methods and shows superior performance at all super-resolution scales on DIV2K and other datasets. Our code will be available at the following locations.  %\url{http://xxxxxxxxxxxxxxxxxxxx}

%%%%%%%%%%%%%%%%
Image representation is critical for many visual tasks. Instead of representing images discretely with 2D arrays of pixels, a recent study, namely local implicit image function (LIIF), denotes images as a continuous function where pixel values are expansion by using the corresponding coordinates as inputs. Due to its continuous nature, LIIF can be adopted for arbitrary-scale image super-resolution tasks, resulting in a single effective and efficient model for various up-scaling factors. However, LIIF often suffers from structural distortions and ringing artifacts around edges, mostly because all pixels share the same model, thus ignoring the local properties of the image. In this paper, we propose a novel adaptive local image function (A-LIIF) to alleviate this problem. Specifically, our A-LIIF consists of two main components: an encoder and a expansion network. The former captures cross-scale image features, while the latter models the continuous up-scaling function by a weighted combination of multiple local implicit image functions. Accordingly, our A-LIIF can reconstruct the high-frequency textures and structures more accurately. Experiments on multiple benchmark datasets verify the effectiveness of our method. Our codes are available at \url{https://github.com/LeeHW-THU/A-LIIF}.

\end{abstract}

\section{Introduction}
\label{sec:intro}
Deep neural networks (DNNs) have been widely and successfully adopted in many applications, such as image super-resolution. Currently, most existing main-stream DNN-based super-resolution methods treat images as discrete 2D arrays and learn an encoder-decoder style model for each up-scaling ratio. As such, developers need to retrain or adopt different networks for different up-scaling ratios. 

\begin{figure}[htp!] %page1效果图
  \centering
   \includegraphics[width=0.98\linewidth]{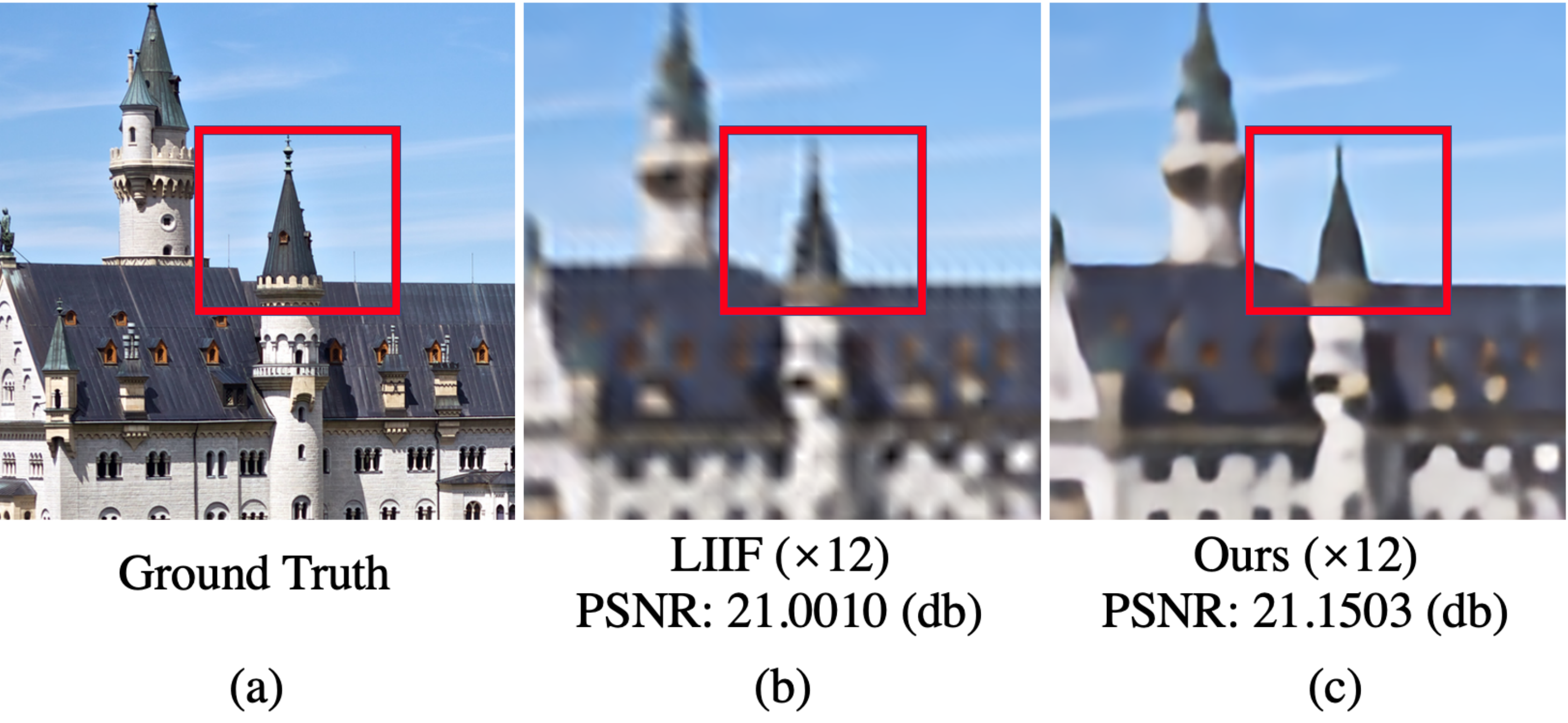}
   \caption{An example of amplified images generated by LIIF \cite{liif} and our A-LIIF at the resolution scale $\times12$. As denoted in the red box areas, LIIF \cite{liif} has ringing artifacts around edges, while our A-LIIF can relieve this problem.
%   are avoided by our A-LIIF.
   }
   \label{fig:p1}
\end{figure}

Recently, the local implicit image function (LIIF) \cite{liif} explores this task from another	perspective, opening the new era of image super-resolution. Specifically, LIIF \cite{liif} shows that high-resolution (HR) images can be learned as continuous functions of low-resolution (LR) features and pixel coordinates, where the function can be easily modeled by a simple multi-layer perception (MLP). Due to its continuous manner, users can exploit LIIF \cite{liif} for arbitrary-scale super-resolution, without creating and training different models.

However, we observe that LIIF \cite{liif} often generates ringing artifacts around edges, especially when the resolution scale is relatively large (as shown in Figure \ref{fig:p1} (b)). This problem is mainly because all predicted pixels share the same MLP whereas pixels from different regions (such as smooth and edge areas)   may have different characteristics. It motivates us to predict pixels by considering the pixel characteristics.

\begin{figure*}[t] %模型图
  \centering
   \includegraphics[width=0.95\linewidth]{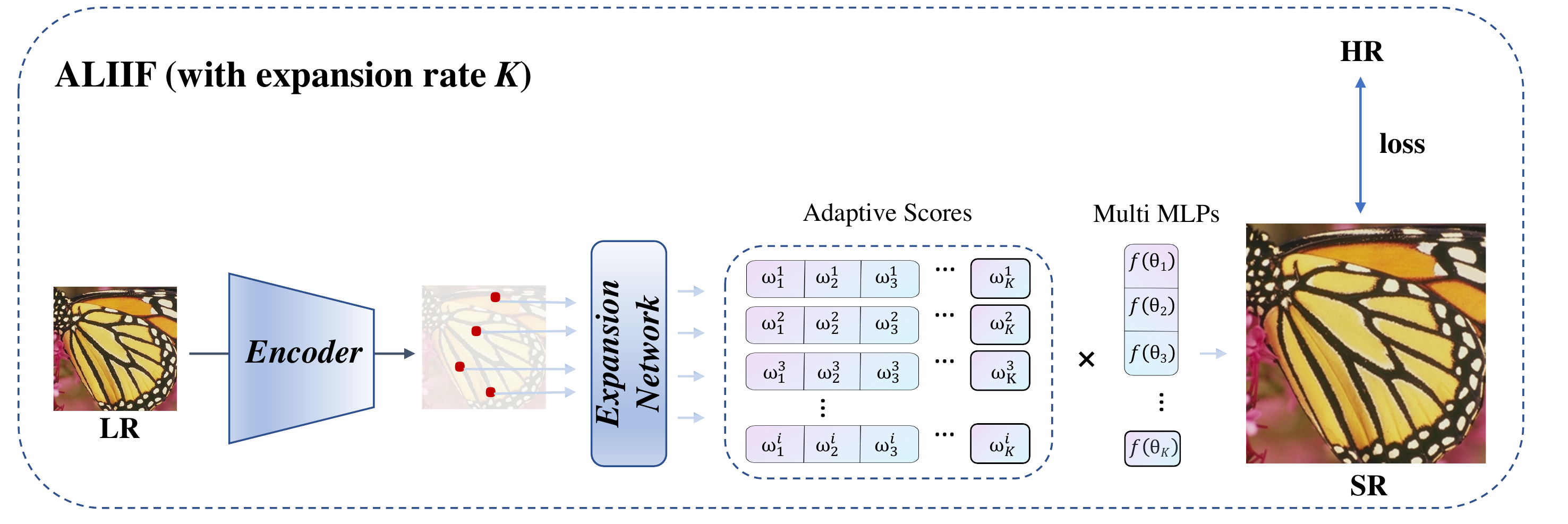}
   \caption{The main pipeline of our A-LIIF. The red dots are the positions of targeted pixels that we need to render.}
   \label{fig:model}
\end{figure*}

In this paper, we propose an adaptive local image function (A-LIIF) by automatically modeling pixel differences with multiple MLPs.  As shown in Figure \ref{fig:model},  our A-LIIF mainly consists of encoder, expansion network, and multiple MLPs.
Specifically, we design an expansion network  to generate $K$ different adaptive scores of each element in the feature map, where $K$ is the expansion rate. After that, the predicted pixel is produced by a weighted linear combination of $K$ MLPs with the adaptive scores. 
Unlike the LIIF \cite{liif} that suffers from ringing artifacts, our A-LIIF can relieve such problem  and reconstruct the details more accurately (seen in Figure \ref{fig:p1}(c)).

\vspace{0.35em}
In conclusion, the main contributions of this paper can be summarized as follows:
\begin{itemize}
    \item We reveal that LIIF \cite{liif} usually suffers from structural distortions and ringing artifacts around edges, mostly because all pixels share the same MLP.
    \item We propose an adaptive local image function (A-LIIF) by modeling pixel differences with multiple MLPs automatically to alleviate artifacts.
    \item We conduct extensive experiments on five benchmark datasets to verify the effectiveness of our A-LIIF.
\end{itemize}

\section{RELATED WORKs}
\label{sec:related}
This section provides a brief overview of recent work on implicit neural representations, expansion network, basis, and various super-resolution methods related to our work.
\subsection{Single Image Super-Resolution}
\label{ssec:SISR}
Super-resolution is one of the low-level vision tasks that has been explored in our vision field for decades, especially single image super-resolution (SISR). With the rapid advancement of deep learning techniques, super-resolution models implemented based on deep learning are also in the process of breakthrough. Different types of deep learning strategies are widely applied to target super-resolution related problems. The earliest were adopted CNN algorithms SRCNN  \cite{srcnn}, and later, networks with larger residual structures [VDSR \cite{vdsr}, IRCNN \cite{ircnn}, SRResNet \cite{srresnet}, EDSR \cite{edsr}, RDN \cite{rdn}] , while the concept of non-localization is also very prevalent [SENet \cite{SENet}, Non-local \cite{non-local}, SAN \cite{san}] taking the 
SR quality to a new level, all these works have actively explored the field of SR.
\subsection{Arbitrary-Scale Super-Resolution}
\label{ssec:ASSR}
Arbitrary-scale super-resolution methods hold great promise for research. The emergence of traditional single-image super-resolution methods, whose models can only be used at a single scale, a phenomenon that makes it difficult to land super-resolution modeling methods commercially in real life, greatly surpasses previous SISR work in terms of convenience. The earliest arbitrary-scale super-resolution methods can be traced back to MetaSR \cite{metasr}, which was the first CNN-based arbitrary-scale SR method. The recent SOTA work LIIF \cite{liif} proposes a new framework for arbitrary-scale super-resolution using implicit neural representations, where feature maps, coordinates, and scaling factors are input to the MLP after Encoder for computation to obtain RGB values.

\begin{table*}[htp!]%DIV2K对比表格
%\footnotesize
\setlength{\tabcolsep}{4mm}
\centering
\caption{Quantitative comparison on DIV2K \cite{DIV2K} validation set (PSNR (dB)). $\phi$ indicates ours implementation. EDSR-baseline \cite{edsr} trains different models for different scales.}
\begin{tabular}{c|ccc|ccccc}
\toprule
\multirow{2}{*}{Method} & \multicolumn{3}{c|}{In-distribution}             & \multicolumn{5}{c}{Out-of-distribution}                                            \\
                        & ×2             & ×3             & ×4             & ×6             & ×12            & ×18            & ×24            & ×30            \\ \hline
bicubic \cite{edsr}                 & 31.01          & 28.22          & 26.66          & 24.82          & 22.27          & 21.00             & 20.19          & 19.59          \\
EDSR-baseline \cite{edsr}           & 34.56          & 30.82          & 28.84          & -              & -              & -              & -              & -              \\ \hline
EDSR-baseline-LIIF \cite{liif}$^{\phi}$     & 33.57          & 29.93          & 28.14          & 26.02          & 23.17          & 21.74          & 20.83          & 20.19          \\
A-LIIF     & \textbf{34.65} & \textbf{30.95} & \textbf{28.99} & \textbf{26.75} & \textbf{23.71} & \textbf{22.18} & \textbf{21.18} & \textbf{20.48} \\ \bottomrule
\end{tabular}

   \label{table:table1}
\end{table*}

\section{The Proposed METHOD}
\label{sec:METHODS}
Our work is inspired by NeRF \cite{nerf}, which demonstrates that 3D scenes with fine textures can be rendered with a neural rendering approach. A recent work named LIIF \cite{liif} further demonstrates that implicit image functions can recover images at arbitrary scales of super-resolution with accurate details. However, the LIIF \cite{liif} approach usually suffers from ringing artifacts by using a shared MLP for every pixel point.
Thus, a natural question arises: does it use the same implicit neural function even for very different edge features? For this purpose, we design a novel adaptive local image function (A-LIIF) to relieve such problem, which mainly consists of expansion network and multiple MLPs modules. Besides,
our method demonstrates that for each pixel point a different MLPs should be used to recover high frequency textures. 
% where the two main methods are the expansion network as well as the Multi MLP module, and the combination of these two methods enables the adaptive expression of the corresponding implicit function for each pixel point. 
The main diagram of our method is shown in Figure \ref{fig:model}, which is detailed in the following sections.

\subsection{Expansion Network}
\label{ssec:PredictNet}
%TODO
The edge features of all pixel points are different when recovering an image, however LIIF \cite{liif} ignores this importance and directly for all pixel points is recovered using a shared implicit image function. We empirically note that these cases of structural distortion and apparent artifacts arising from neural representations can be resolved if an adaptive MLP layer is used. So our A-LIIF feeds the output of backbone into the expansion network to adaptively expansion $K$ (\ref{predict_eq}).
\begin{equation}
\omega_{K}^{i}=N(P(\tau, \xi)), \quad \xi\propto x-x_r. \label{predict_eq}
\end{equation}
%Then the received signal \eqref{XX}
where $\tau$ refers to the feature map output by backbone, $\xi$ refers to the normalized distance between the target pixel location $x$ and the reference feature location $x_r$, $P$ refers to our expansion network, and $N$ refers to the fact that we normalize the output of the expansion network so that the sum is guaranteed to be $1$. So the final output $\omega_{K}^{i}$ refers to the $K$ expansion values of the current $i$-th pixel point. Here our expansion network uses a 5-layer MLP for expansion.

\subsection{Multiple MLPs}
\label{ssec:MultiMLP}
The expansion network generates $K$ different adaptive scores of each element in the feature map. After that, the predicted pixel is produced by a weighted linear combination of $K$ MLPs with the adaptive scores. This is combined with the $K$ designed basis, and this fusion ensures that all pixel points are the result of the adaptive combination of the $K$ basis (\ref{mlp_eq}).
\begin{equation}
SR=ReLU(\sum_{K}^{} \tau\cdot (\omega _{K}^{i}\times f(\theta_K))). \label{mlp_eq}
\end{equation}

\section{EXPERIMENTS}
\label{sec:EXPERIMENT}
In this section, we describe the training details and dataset used for our experiments. We then demonstrate the performance of our A-LIIF by comparing it with other models.
\subsection{Dataset}
\label{ssec:Dataset}
We use the DIV2K \cite{DIV2K} dataset from the NTIRE 2017 challenge to train and evaluate the model. In this, we use 800 images from the DIV2K \cite{DIV2K} training set for training and 100 images from the validation set for testing. Following much of the previous work, Our model also compares performance on the [Set5 \cite{set5}, Set14 \cite{set14}, B100 \cite{b100}, Urban100 \cite{u100}] benchmark dataset. Finally, we use the widely adopted PSNR as an evaluation metric for the model.

\begin{table*}[htp!]%benchmark对比表格
%\footnotesize
\setlength{\tabcolsep}{5mm}
\centering
\caption{Quantitative comparison on benchmark datasets (PSNR (dB)). $\phi$ indicates ours implementation. EDSR-baseline \cite{edsr} trains different models for different scales.}
\begin{tabular}{c|c|ccc|cc}
\toprule
\multirow{2}{*}{Dataset}  & \multirow{2}{*}{Method} & \multicolumn{3}{c|}{In-distribution}             & \multicolumn{2}{c}{Out-of-distribution} \\
                          &                         & ×2             & ×3             & 4              & ×6                 & ×8                 \\ \hline
\multirow{3}{*}{Set5 \cite{set5}}     & EDSR-baseline \cite{edsr}           & 34.99          & 31.07          & 29.14          & -                  & -                  \\
                          & EDSR-baseline-LIIF \cite{liif}$^{\phi}$     & 36.99          & 33.08          & 30.91          & 27.83              & 26.1               \\
                          & A-LIIF (ours)            & \textbf{37.96} & \textbf{34.39} & \textbf{32.19} & \textbf{28.89}     & \textbf{26.97}     \\ \hline
\multirow{3}{*}{Set14 \cite{set14}}    & EDSR-baseline \cite{edsr}           & 31.41          & 28.21          & 26.59          & -                  & -                  \\
                          & EDSR-baseline-LIIF \cite{liif}$^{\phi}$     & 32.73          & 29.47          & 27.78          & 25.65              & 24.32              \\
                          & A-LIIF (ours)            & \textbf{33.64} & \textbf{30.36}  & \textbf{28.65} & \textbf{26.47}     & \textbf{24.92}     \\ \hline
\multirow{3}{*}{B100 \cite{b100}}     & EDSR-baseline \cite{edsr}           & 30.47          & 27.64          & 26.31          & -                  & -                  \\
                          & EDSR-baseline-LIIF \cite{liif}$^{\phi}$     & 31.47          & 28.49          & 27.05          & 25.40               & 24.41              \\
                          & A-LIIF (ours)            & \textbf{32.18} & \textbf{29.11} & \textbf{27.60} & \textbf{25.84}     & \textbf{24.80}     \\ \hline
\multirow{3}{*}{Urban100 \cite{u100}} & EDSR-baseline \cite{edsr}           & 27.91          & 25.00             & 23.59          & -                  & -                  \\
                          & EDSR-baseline-LIIF \cite{liif}$^{\phi}$     & 29.98          & 26.46          & 24.74          & 22.82              & 21.71              \\
                          & A-LIIF (ours)            & \textbf{32.09} & \textbf{28.19} & \textbf{26.14} & \textbf{23.80}     & \textbf{22.47}    \\ \bottomrule
\end{tabular}

   \label{table:table2}
\end{table*}

\subsection{Training Details}
\label{ssec:Training Details}
In training A-LIIF, we follow the setup of the LIIF \cite{liif} model. We uniformly down-sampled the GT HR images from the training set of 800 images in DIV2K \cite{DIV2K} by bicubic \cite{edsr} interpolation (x2$\sim$x4) as our LR training set. For the input model, we randomly crop the LR images into 48x48 patches, collect 2304 random pixels on the HR images, and use L1 loss to invert the ground truth. For the choice of optimizer, we use an ADAM optimizer like LIIF \cite{liif}. The initial learning rate is set to 10-4 and decays by half every 200 calendars. We will train for 10-3 calendar times and perform 10-3 iterations per calendar time. For the Encoder selection, we use EDSR \cite{edsr} as our Encoder. for the design of the expansion network, we use a 5-layer 256-neuron MLP with $K$ outputs of 10 expansion values. For the multi-MLP design, we use a 5-layer 16-neuron MLP for each pixel point. we Since our model has 16 neurons for each pixel, we set the number of neurons in the later MLP layers to 16 when replicating the LIIF \cite{liif}.

\begin{figure}[!t] %Ablation Studies效果图
  \centering
   \includegraphics[width=0.95\linewidth]{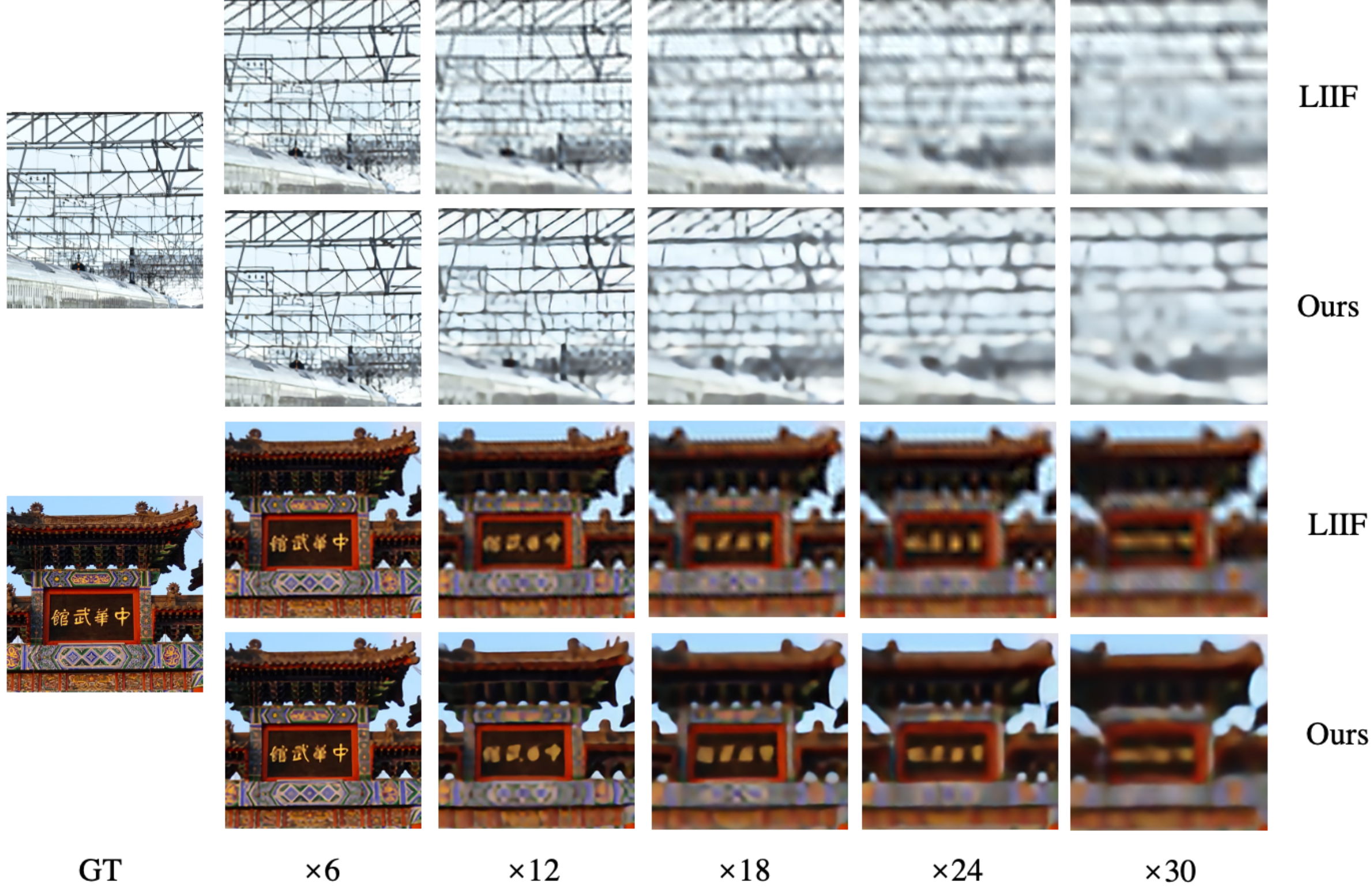}
   \includegraphics[width=0.95\linewidth]{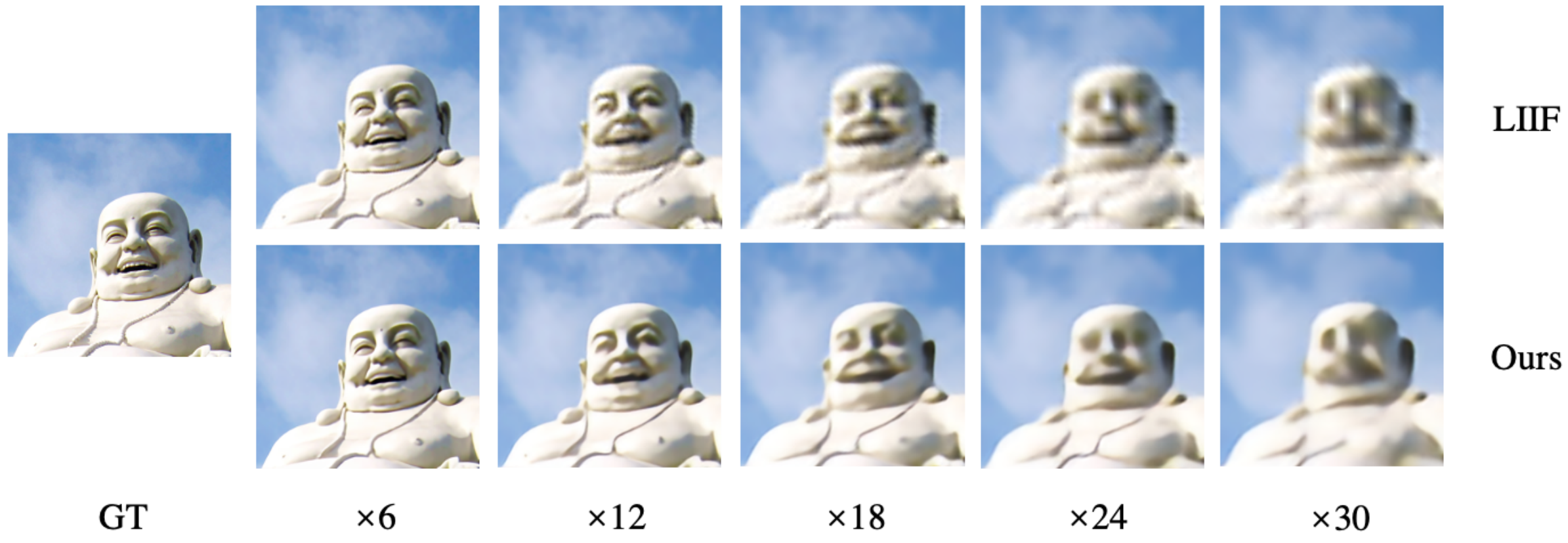}
   \caption{Qualitative comparison between LIIF \cite{liif} and A-LIIF.}
   \label{fig:res1}
\end{figure}

\subsection{Main Results}
\label{ssec:Result}
% hack: manually set the reference number
% \manuallabel{fig:res2}{4}
% \manuallabel{fig:res1}{5}
Table \ref{table:table1} compares the performance of LIIF \cite{liif} and our A-LIIF at different resolution scales on DIV2K \cite{DIV2K}. We note that the PSNR results of our model are consistently higher than those of LIIF \cite{liif} at all scales.

We use the red box areas in Figure \ref{fig:p1} as an example to compare LIIF and our methods via SSIM and PSNR. The SSIM of LIIF and our method are 0.8572 and 0.8802, respectively. The PSNR of LIIF and our method are 22.1157 and 22.8543, respectively. These results quantitatively verify that LIIF usually suffers from distortions or ringing artifacts around edges whereas our method is not.

In Table \ref{table:table2}, we also compare A-LIIF with other methods on fore standard benchmark datasets: Set5 \cite{set5}, Set14 \cite{set14}, B100 \cite{b100} and  Urban100 \cite{u100}. We again demonstrate the superiority of our A-LIIF, and our model outperforms both EDSR \cite{edsr} and LIIF \cite{liif} in all experiments, even though EDSR \cite{edsr} trains a dedicated model for each scale.
Finally, we show a qualitative comparison between LIIF \cite{liif} and A-LIIF. as shown in Figure \ref{fig:res1}. Our A-LIIF avoids the effect of LIIF \cite{liif} structure loss with high frequency textures by using a fusion of expansion network and basis for each pixel. 

We also compare our method on RDN \cite{rdn} backbone. We only train models 100 epochs instead of 1,000 epochs. The PSNR of RDN-LIIF, and RDN-A-LIIF (Ours) are 27.94 and 28.73, respectively. These results show that our method is still better than LIIF \cite{liif}.

\subsection{The Ablation Study}
\label{ssec:Ablation Studies}
\begin{figure}[!t] %Ablation Studies效果图
  \centering
   \includegraphics[width=0.9\linewidth]{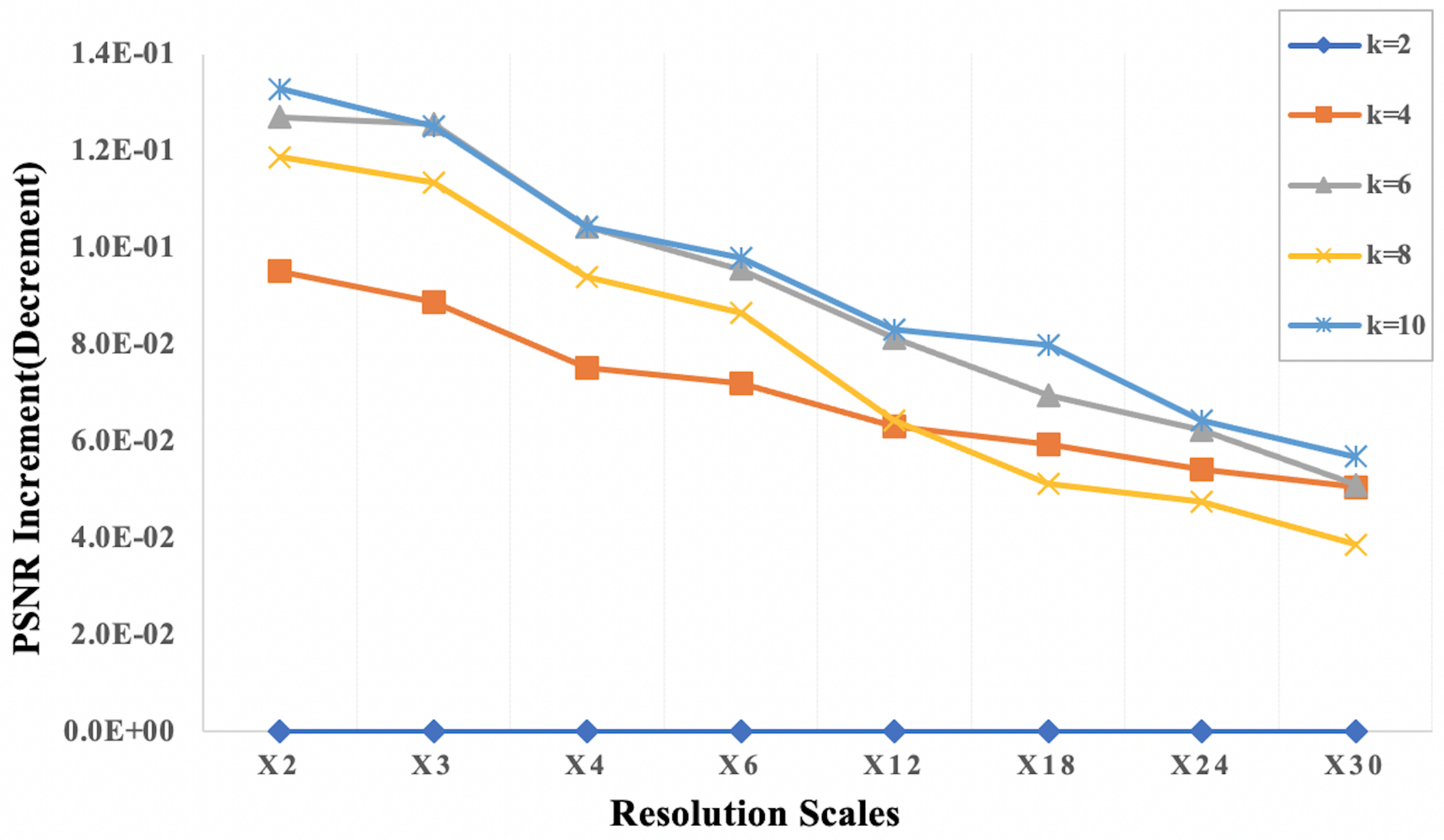}
   \caption{The choice of hyper-parameter $K$.}
   \label{fig:Ablation}
\end{figure}

In our ablation study, we trained several models in which we gradually added our new design and explored the choice of hyper-parameter $K$. For a fair comparison, we used EDSR \cite{edsr} as our encoder in all models. The training setup was kept constant, where the same ADAM optimizer, learning rate, scheduler, and training length were used. The overall results are shown in Fig \ref{fig:Ablation}, Table \ref{table:table1}-\ref{table:table2} demonstrates that using a shared MLP layer for all pixel points for recovery does not work well, but using an adaptive MLP for each pixel point leads to promising results.

\section{Conclusion}
\label{sec:Conclusion}

In this paper, we introduced a novel arbitrary-scale super-resolution model, dubbed A-LIIF, that combined implicit image functions with an expansion network and the depth of multiple MLPs. It revealed the importance of using different MLPs for different pixel points. We also verified the importance of using non-shared MLPs by experiments.

% \section*{Acknowledgments}
% This work is supported in part by the National Natural Science Foundation of China under Grant 62171248 and the PCNL Key Project under Grant PCL2021A07.

\vfill\pagebreak %结束
\clearpage
\bibliographystyle{IEEEbib}
\bibliography{strings,refs}

\begin{thebibliography}{10}

\bibitem{liif}
Yinbo Chen, Sifei Liu, and Xiaolong Wang,
\newblock ``Learning continuous image representation with local implicit image
  function,''
\newblock in {\em Proceedings of the IEEE/CVF Conference on Computer Vision and
  Pattern Recognition}, 2021, pp. 8628--8638.

\bibitem{srcnn}
Chao Dong, Chen~Change Loy, Kaiming He, and Xiaoou Tang,
\newblock ``Image super-resolution using deep convolutional networks,''
\newblock {\em IEEE transactions on pattern analysis and machine intelligence},
  vol. 38, no. 2, pp. 295--307, 2015.

\bibitem{vdsr}
Jiwon Kim, Jung~Kwon Lee, and Kyoung~Mu Lee,
\newblock ``Accurate image super-resolution using very deep convolutional
  networks,''
\newblock in {\em Proceedings of the IEEE conference on computer vision and
  pattern recognition}, 2016, pp. 1646--1654.

\bibitem{ircnn}
Kai Zhang, Wangmeng Zuo, Shuhang Gu, and Lei Zhang,
\newblock ``Learning deep cnn denoiser prior for image restoration,''
\newblock in {\em Proceedings of the IEEE conference on computer vision and
  pattern recognition}, 2017, pp. 3929--3938.

\bibitem{srresnet}
Christian Ledig, Lucas Theis, Ferenc Husz{\'a}r, Jose Caballero, Andrew
  Cunningham, Alejandro Acosta, Andrew Aitken, Alykhan Tejani, Johannes Totz,
  Zehan Wang, et~al.,
\newblock ``Photo-realistic single image super-resolution using a generative
  adversarial network,''
\newblock in {\em Proceedings of the IEEE conference on computer vision and
  pattern recognition}, 2017, pp. 4681--4690.

\bibitem{edsr}
Bee Lim, Sanghyun Son, Heewon Kim, Seungjun Nah, and Kyoung Mu~Lee,
\newblock ``Enhanced deep residual networks for single image
  super-resolution,''
\newblock in {\em Proceedings of the IEEE conference on computer vision and
  pattern recognition workshops}, 2017, pp. 136--144.

\bibitem{rdn}
Yulun Zhang, Yapeng Tian, Yu~Kong, Bineng Zhong, and Yun Fu,
\newblock ``Residual dense network for image super-resolution,''
\newblock in {\em Proceedings of the IEEE conference on computer vision and
  pattern recognition}, 2018, pp. 2472--2481.

\bibitem{SENet}
Jie Hu, Li~Shen, and Gang Sun,
\newblock ``Squeeze-and-excitation networks,''
\newblock in {\em Proceedings of the IEEE conference on computer vision and
  pattern recognition}, 2018, pp. 7132--7141.

\bibitem{non-local}
Xiaolong Wang, Ross Girshick, Abhinav Gupta, and Kaiming He,
\newblock ``Non-local neural networks,''
\newblock in {\em Proceedings of the IEEE conference on computer vision and
  pattern recognition}, 2018, pp. 7794--7803.

\bibitem{san}
Tao Dai, Jianrui Cai, Yongbing Zhang, Shu-Tao Xia, and Lei Zhang,
\newblock ``Second-order attention network for single image super-resolution,''
\newblock in {\em Proceedings of the IEEE/CVF conference on computer vision and
  pattern recognition}, 2019, pp. 11065--11074.

\bibitem{metasr}
Xuecai Hu, Haoyuan Mu, Xiangyu Zhang, Zilei Wang, Tieniu Tan, and Jian Sun,
\newblock ``Meta-sr: A magnification-arbitrary network for super-resolution,''
\newblock in {\em Proceedings of the IEEE/CVF Conference on Computer Vision and
  Pattern Recognition}, 2019, pp. 1575--1584.

\bibitem{DIV2K}
Eirikur Agustsson and Radu Timofte,
\newblock ``Ntire 2017 challenge on single image super-resolution: Dataset and
  study,''
\newblock in {\em Proceedings of the IEEE conference on computer vision and
  pattern recognition workshops}, 2017, pp. 126--135.

\bibitem{nerf}
Ben Mildenhall, Pratul~P Srinivasan, Matthew Tancik, Jonathan~T Barron, Ravi
  Ramamoorthi, and Ren Ng,
\newblock ``Nerf: Representing scenes as neural radiance fields for view
  synthesis,''
\newblock in {\em European conference on computer vision}. Springer, 2020, pp.
  405--421.

\bibitem{set5}
Marco Bevilacqua, Aline Roumy, Christine Guillemot, and Marie~Line
  Alberi-Morel,
\newblock ``Low-complexity single-image super-resolution based on nonnegative
  neighbor embedding,''
\newblock 2012.

\bibitem{set14}
Roman Zeyde, Michael Elad, and Matan Protter,
\newblock ``On single image scale-up using sparse-representations,''
\newblock in {\em International conference on curves and surfaces}. Springer,
  2010, pp. 711--730.

\bibitem{b100}
David Martin, Charless Fowlkes, Doron Tal, and Jitendra Malik,
\newblock ``A database of human segmented natural images and its application to
  evaluating segmentation algorithms and measuring ecological statistics,''
\newblock in {\em Proceedings Eighth IEEE International Conference on Computer
  Vision. ICCV 2001}. IEEE, 2001, vol.~2, pp. 416--423.

\bibitem{u100}
Jia-Bin Huang, Abhishek Singh, and Narendra Ahuja,
\newblock ``Single image super-resolution from transformed self-exemplars,''
\newblock in {\em Proceedings of the IEEE conference on computer vision and
  pattern recognition}, 2015, pp. 5197--5206.

\end{thebibliography}

\end{document}